\documentclass[prl,preprint,amsmath,amssymb]{revtex4}
\usepackage{graphicx}
\usepackage[justification=centering]{caption}

\begin{document}

\title{Tighter monogamy relations of quantum entanglement for multiqubit W-class states}

\author{Zhi-Xiang Jin$^{1}$}
\author{Shao-Ming Fei$^{1,2}$}

\affiliation{$^1$School of Mathematical Sciences, Capital Normal University,
Beijing 100048, China\\
$^2$Max-Planck-Institute for Mathematics in the Sciences, 04103 Leipzig, Germany}

\bigskip

\begin{abstract}

Monogamy relations characterize the distributions of entanglement in multipartite systems. We investigate monogamy relations for multiqubit generalized $W$-class states. We present new analytical monogamy inequalities for the concurrence of assistance, which are shown to be tighter than the existing ones. Furthermore, analytical monogamy inequalities are obtained for the negativity of assistance.

\end{abstract}

\maketitle

\section{INTRODUCTION}
Quantum entanglement \cite{MAN,RPMK,FMA,KSS,HPB,HPBO,JIV,CYSG} is an essential feature of quantum mechanics. As one of the fundamental differences between quantum entanglement and classical correlations, a key property of entanglement is that a quantum system entangled with one of other subsystems limits its entanglement with the remaining ones. The monogamy relations give rise to the distribution of entanglement in the multipartite setting. Monogamy is also an essential feature allowing for security in quantum key distribution \cite{MP}.	
  									
For a tripartite system $A$, $B$ and $C$, the usual monogamy of an entanglement measure $\mathcal{E}$ implies that \cite{MK} the entanglement between $A$ and $BC$ satisfies $\mathcal{E}_{A|BC}\geq \mathcal{E}_{AB} +\mathcal{E}_{AC}$. In Ref. \cite{KSJ,KSJB}, the monogamy of entanglement for multiqubit $W$-class states has been investigated, and the monogamy relations for tangle and the squared concurrence have been proved. It gives the general monogamy relations for the $x$-power \cite{ZXN} of concurrence of assistance for generalized multiqubit $W$-class states.

In this paper, we show that the monogamy inequalities for concurrence of assistance obtained so far can be made tighter. We establish entanglement monogamy relations for the $x$-th $(x\geq2)$ and $y$-th $(y<0)$ power of the concurrence of assistance which are tighter than those in \cite{ZXN}, which give rise to finer characterizations of the entanglement distributions among the multipartite $W$-class states. Furthermore, we also present the general monogamy relations for the $x$-power of negitivity of assistance for generalized multiqubit $W$-class states.

\section{TIGHTER MONOGAMY RELATIONS FOR CONCURRENCE OF ASSISTANCE}

We first consider the monogamy inequalities related to concurrence. Let $H_X$ denote a discrete finite dimensional complex vector space associated with a quantum subsystem $X$.
For a bipartite pure state $|\psi\rangle_{AB}$ in vector space $H_A\otimes H_B$, the concurrence is given by \cite{AU,PR,SA}
\begin{equation}\label{CD}
C(|\psi\rangle_{AB})=\sqrt{{2\left[1-\mathrm{Tr}(\rho_A^2)\right]}},
\end{equation}
where $\rho_A$ is the reduced density matrix by tracing over the subsystem $B$, $\rho_A=\mathrm{Tr}_B(|\psi\rangle_{AB}\langle\psi|)$. The concurrence for a bipartite mixed state $\rho_{AB}$ is defined by the convex roof extension
\begin{equation*}
 C(\rho_{AB})=\min_{\{p_i,|\psi_i\rangle\}}\sum_ip_iC(|\psi_i\rangle),
\end{equation*}
where the minimum is taken over all possible decompositions of $\rho_{AB}=\sum_ip_i|\psi_i\rangle\langle\psi_i|$, with $p_i\geq0$ and $\sum_ip_i=1$ and $|\psi_i\rangle\in H_A\otimes H_B$.

For a tripartite state $|\psi\rangle_{ABC}$, the concurrence of assistance is defined by \cite{TFS, YCS}
\begin{eqnarray*}
C_a(|\psi\rangle_{ABC})\equiv C_a(\rho_{AB})=\mathrm{max}_{\{p_i,|\psi_i\rangle\}}\sum_ip_iC(|\psi_i\rangle),
\end{eqnarray*}
where the maximum is taken over all possible decompositions of $\rho_{AB}=\mathrm{Tr}_C(|\psi\rangle_{ABC}\langle\psi|)=\sum_ip_i|\psi_i\rangle_{AB}\langle\psi_i|.$ When $\rho_{AB}=|\psi\rangle_{AB}\langle\psi|$ is a pure state, then one has $C(|\psi\rangle_{AB})=C_a(\rho_{AB})$.

For an $N$-qubit pure state $|\psi\rangle_{AB_1\cdots B_{N-1}}\in H_A\otimes H_{B_1}\otimes\cdots\otimes H_{B_{N-1}}$, the concurrence $C(|\psi\rangle_{A|B_1\cdots B_{N-1}})$ of the state $|\psi\rangle_{A|B_1\cdots B_{N-1}}$, viewed as a bipartite state under the partition $A$ and $B_1,B_2,\cdots, B_{N-1}$, satisfies \cite{ZF}
\begin{equation*}\label{CA}
 C^{\alpha}(\rho_{A|B_1,B_2\cdots,B_{N-1}})\geq C^{\alpha}(\rho_{AB_1})+C^{\alpha}(\rho_{AB_2})+\cdots+C^{\alpha}(\rho_{AB_{N-1}}),
\end{equation*}
for $\alpha\geq2$, where $\rho_{AB_i}=\mathrm{Tr}_{B_1\cdots B_{i-1}B_{i+1}\cdots B_{N-1}}(|\psi\rangle_{AB_1\cdots B_{N-1}}\langle\psi|)$.
It is further improved that for $\alpha\geq2$, one has \cite{JZX},
\begin{eqnarray}\label{l1}
C^\alpha(\rho_{A|B_1B_2\cdots B_{N-1}})&&\geq C^\alpha(\rho_{AB_1})
 +\frac{\alpha}{2} C^\alpha(\rho_{AB_2})+\cdots+\left(\frac{\alpha}{2}\right)^{m-1}C^\alpha(\rho_{AB_m})\\ \nonumber
 &&+\left(\frac{\alpha}{2}\right)^{m+1}\left(C^\alpha(\rho_{AB_{m+1}})
 +\cdots+C^\alpha(\rho_{AB_{N-2}})\right)
+\left(\frac{\alpha}{2}\right)^{m}C^\alpha(\rho_{AB_{N-1}})
\end{eqnarray}
and
\begin{equation}\label{l2}
  C^\alpha(\rho_{A|B_1B_2\cdots B_{N-1}})< K\left(C^\alpha(\rho_{AB_1})+C^\alpha(\rho_{AB_2})+\cdots+C^\alpha(\rho_{AB_{N-1}})\right)
\end{equation}
for all $\alpha<0$, where $K=\frac{1}{N-1}$.

Dual to the Coffman-Kundu-Wootters inequality, the generalized monogamy relation based on the concurrence of assistance do not satisfy the monogamy relation. But, for an  $N$-qubit generlized $W$-class states $|\psi\rangle_{AB_1\cdots B_{N-1}}\in H_A\otimes H_{B_1}\otimes\cdots\otimes H_{B_{N-1}}$, the concurrence of assistance $C_a(|\psi\rangle_{A|B_1\cdots B_{N-1}})$ of the state $|\psi\rangle_{AB_1\cdots B_{N-1}}$ satisfies the inequality \cite{ZXN},
\begin{equation}\label{ZXN1}
  C_a^x(\rho_{A|B_1,B_2\cdots,B_{N-1}})\geq C_a^x(\rho_{AB_1})+C_a^x(\rho_{AB_2})+\cdots+C_a^x(\rho_{AB_{N-1}}),
\end{equation}
and
\begin{equation}\label{ZXN2}
  C_a^y(\rho_{A|B_1,B_2\cdots,B_{N-1}})< C_a^y(\rho_{AB_1})+C_a^y(\rho_{AB_2})+\cdots+C_a^y(\rho_{AB_{N-1}}),
\end{equation}
where $x\geq2$, $y\leq0$.

In fact, as the characterization of the entanglement distribution among the subsystems, the monogamy inequalities satisfied by the concurrence of assistance can be further refined and become tighter.

In the following, we study the monogamy property of the concurrence of assistance for the $N$-qubit generalized $W$-class states $|\psi\rangle\in H_A\otimes H_{B_1}\otimes\cdots\otimes H_{B_{N-1}}$ defined by
\begin{eqnarray}\label{gw}
|\psi\rangle=a|00\cdots0\rangle+b_1|10\cdots0\rangle+\cdots+b_N|00\cdots1\rangle,
\end{eqnarray}
with $|a|^2+\sum_{i=1}^N|b_i|^2=1$. For the $N$-qubit generalized $W$-class states (\ref{gw}), one has \cite{ZXN},
\begin{eqnarray}\label{la2}
C(\rho_{AB_i})=C_a(\rho_{AB_i}),~~~~i=1,2,...,N-1,
\end{eqnarray}
where $\rho_{AB_i}=\mathrm{Tr}_{B_1\cdots B_{i-1}B_{i+1}\cdots B_{N-1}}(|\psi\rangle\langle\psi|)$.

{[\bf Theorem 1]}.
For the $N$-qubit generalized $W$-class states $|\psi\rangle\in H_A\otimes H_{B_1}\otimes\cdots\otimes H_{B_{N-1}}$, let $\rho_{AB_{j_1}\cdots B_{j_{m-1}}}$ denote the $m$-qubit, $2\leq m\leq N$, reduced density matrix of $|\psi\rangle$. If $C(\rho_{AB_{j_i}})\geq C(\rho_{AB_{j_{i+1}}\cdots B_{j_{m-1}}})$ for $i=1,2,\cdots t$, and $C(\rho_{AB_{j_k}})\leq C(\rho_{AB_{j_{k+1}}\cdots B_{j_{m-1}}})$ for $k=t+1,\cdots,{m-2}$, $\forall$ $1\leq t\leq {m-3},~m\geq 4$, the concurrence of assistance satisfies
\begin{eqnarray}\label{th1}
&&C_a^x(\rho_{A|B_{j_1}\cdots B_{j_{m-1}}})\geq C_a^x(\rho_{AB_{j_1}}) \nonumber\\
&&+\frac{x}{2}C_a^x(\rho_{AB_{j_2}})+\cdots+\left(\frac{x}{2}\right)^{t-1}C_a^x(\rho_{AB_{j_t}}) \nonumber\\
&&+\left(\frac{x}{2}\right)^{t+1}\left(C_a^x(\rho_{AB_{j_{t+1}}})+\cdots+C_a^x(\rho_{AB_{j_{m-2}}})\right)\nonumber\\
&&+\left(\frac{x}{2}\right)^tC_a^x(\rho_{AB_{j_{m-1}}})
\end{eqnarray}
for all $x\geq2$.

{\sf [Proof].}
For the $N$-qubit generalized $W$-class states $|\psi\rangle$, according to the definitions of $C(\rho)$ and $C_a(\rho)$, one has $C_a(\rho_{A|B_{j_1}\cdots B_{j_{m-1}}})\geq C(\rho_{A|B_{j_1}\cdots B_{j_{m-1}}})$. When $x\geq2$, we have
\begin{eqnarray}\label{pf1}
C_a^x(\rho_{A|B_{j_1}\cdots B_{j_{m-1}}})
&&\geq C^x(\rho_{A|B_{j_1}\cdots B_{j_{m-1}}}) \geq C^x(\rho_{AB_{j_1}}) \nonumber\\
&&+\frac{x}{2}C^x(\rho_{AB_{j_2}})+\cdots+\left(\frac{x}{2}\right)^{t-1}C^x(\rho_{AB_{j_t}}) \nonumber\\
&&+\left(\frac{x}{2}\right)^{t+1}\left(C^x(\rho_{AB_{j_{t+1}}})+\cdots+C^x(\rho_{AB_{j_{m-2}}})\right)\nonumber\\
&&+\left(\frac{x}{2}\right)^tC^x(\rho_{AB_{j_{m-1}}})\nonumber\\
&&=C_a^x(\rho_{AB_{j_1}})
+\frac{x}{2}C_a^x(\rho_{AB_{j_2}})+\cdots+\left(\frac{x}{2}\right)^{t-1}C_a^x(\rho_{AB_{j_t}}) \nonumber\\
&&+\left(\frac{x}{2}\right)^{t+1}\left(C_a^x(\rho_{AB_{j_{t+1}}})+\cdots+C_a^x(\rho_{AB_{j_{m-2}}})\right)\nonumber\\
&&+\left(\frac{x}{2}\right)^tC_a^x(\rho_{AB_{j_{m-1}}}),
\end{eqnarray}
where we have used in the first inequality the relation $a^x\geq b^x$ for $a\geq b\geq 0,~x\geq 2$. The second inequality is due to (\ref{l1}). The equality is due to (\ref{la2}). \hfill \rule{1ex}{1ex}

As for $x\geq 2$, $(x/2)^t\geq 1$ for all $1\leq t\leq j_{m-3}$, comparing with
the monogamy relations for concurrence of assistance (\ref{ZXN1}), our formula (\ref{th1}) in Theorem 1 gives a tighter monogamy relation with larger lower bounds. In Theorem 1 we have assumed that
some $C(\rho_{AB_{j_i}})\geq C(\rho_{AB_{j_{i+1}}\cdots B_{j_{m-1}}})$ and some
$C(\rho_{AB_k})\leq C(\rho_{AB_{k+1}\cdots B_{m-1}})$ for the $N$-qubit generalized $W$-class states.
If all $C(\rho_{AB_{j_i}})\geq C(\rho_{AB_{j_{i+1}}\cdots B_{j_{m-1}}})$ for $i=1, 2, \cdots, {m-2}$, then we have
the following conclusion:

{\bf [Theorem 2]}.
If $C(\rho_{AB_{j_i}})\geq C(\rho_{AB_{j_{i+1}}\cdots B_{j_{m-1}}})$ for $i=1, 2, \cdots, {m-2}$, then we have
\begin{eqnarray}\label{th2}
C_a^x(\rho_{A|B_{j_1}\cdots B_{j_{m-1}}})\geq C_a^x(\rho_{AB_{j_1}}) +\frac{x}{2}C_a^x(\rho_{AB_{j_2}})+\cdots+\left(\frac{x}{2}\right)^{m-2}C_a^x(\rho_{AB_{j_{m-1}}})
\end{eqnarray}
for all $x\geq2$.

{\it Example 1}. Let us consider the 4-qubit generlized $W$-class states,
\begin{eqnarray}\label{W4}
|W\rangle_{AB_1B_2B_3}=\frac{1}{2}(|1000\rangle+|0100\rangle+|0010\rangle+|0001\rangle).
\end{eqnarray}
We have $C_a^x(|\psi\rangle_{A|B_1B_2B_3})=(\frac{\sqrt{3}}{2})^x$. From our result (\ref{th1}) we have $C_a^x(|\psi\rangle_{A|B_1B_2B_3})\geq \left[1+\frac{x}{2}+(\frac{x}{2})^2\right](\frac{1}{2})^x$, and from (\ref{ZXN1}) one has $C_a^x(|\psi\rangle_{A|B_1B_2B_3})\geq 3(\frac{1}{2})^x$, $x\geq2$. One can see that our result is better than that in \cite{ZXN} for $x\geq2$, see Fig. 1.

\begin{figure}
  \centering
  \includegraphics[width=7cm]{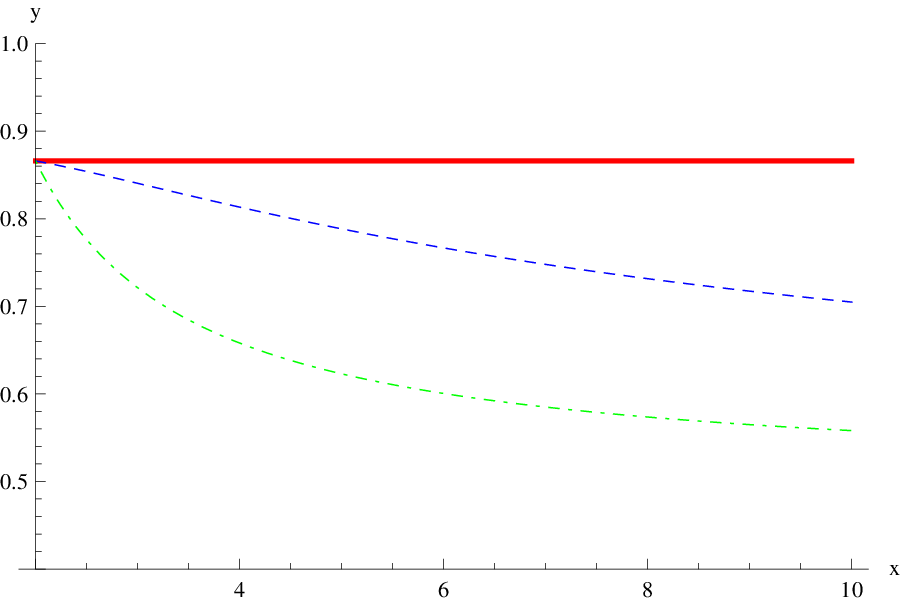}\\
  \caption{$y$ is the value of $C_a(|\psi\rangle_{A|B_1B_2B_3})$. Solid (red) line is the exact value of $C_a(|\psi\rangle_{A|B_1B_2B_3})$, dashed (blue) line is the lower bound of $C_a(|\psi\rangle_{A|B_1B_2B_3})$ in (\ref{th1}), and dot-dashed (green) line is the lower bound in \cite{ZXN} for $x\geq2$.}\label{1}
\end{figure}

We can also derive a tighter upper bound of $C_a^y(\rho_{A|B_1\cdots B_{N-1}})$ for $y<0$.

{[\bf Theorem 3]}.
For the $N$-qubit generalized $W$-class states $|\psi\rangle\in H_A\otimes H_{B_1}\otimes\cdots\otimes H_{B_{N-1}}$, let $\rho_{AB_{j_1}\cdots B_{j_{m-1}}}$ be the $m$-qubit, $2\leq m\leq N$, reduced density matrix of $|\psi\rangle$ with $C(\rho_{AB_{j_i}})\neq 0$ for $1\leq i\leq m-1$, we have
\begin{eqnarray}\label{th3}
  C_a^y(\rho_{A|B_{j_1}\cdots B_{j_{m-1}}})< \tilde{M} \left(C_a^y(\rho_{AB_{j_1}})+C_a^y(\rho_{AB_{j_2}})+\cdots+C_a^y(\rho_{AB_{j_{m-1}}})\right)
\end{eqnarray}
for all $y<0$, where $\tilde{M}=\frac{1}{m-1}$.

{\sf [Proof].}
For $y< 0$, we have
\begin{eqnarray}\label{pf31}
  C_a^y(\rho_{A|B_{j_1}\cdots B_{j_{m-1}}})&&\leq C^y(\rho_{A|B_{j_1}\cdots B_{j_{m-1}}}) \nonumber\\
&& <\tilde{M} \left(C^y(\rho_{AB_{j_1}})+C^y(\rho_{AB_{j_2}})+\cdots+C^y(\rho_{AB_{j_{m-1}}})\right)\nonumber \\
&&= \tilde{M} \left(C_a^y(\rho_{AB_{j_1}})+C_a^y(\rho_{AB_{j_2}})+\cdots+C_a^y(\rho_{AB_{j_{m-1}}})\right),
\end{eqnarray}
where we have used in the first inequality the relation $a^x\leq b^x$ for $a\geq b\geq 0,~x\leq 0$. The second inequality is due to (\ref{l2}). The equality is due to (\ref{la2}). \hfill \rule{1ex}{1ex}

As the factor $\tilde{M}=\frac{1}{m-1}$ is less than one, the inequality (\ref{th3}) is tighter than the one in \cite{ZXN}. This factor $\tilde{M}$ depends on the number of partite $N$. Namely, for larger multipartite systems, the inequality (\ref{th3}) gets even tighter than the one in \cite{ZXN}.

{\it Example 2}. Let us consider again the 4-qubit generlized $W$-class states (\ref{W4}). We have $C_a^y(|\psi\rangle_{A|B_1B_2B_3})=(\frac{\sqrt{3}}{2})^y$. From our result (\ref{th3}) we have $C_a^y(|\psi\rangle_{A|B_1B_2B_3})\leq (\frac{1}{2})^y$, while from (\ref{ZXN2}) one gets $C_a^y(|\psi\rangle_{A|B_1B_2B_3})\leq 3(\frac{1}{2})^y$. It can be seen that our result is better than that in \cite{ZXN} for $y<0$, see Fig. 2.

\begin{figure}
  \centering
  \includegraphics[width=7cm]{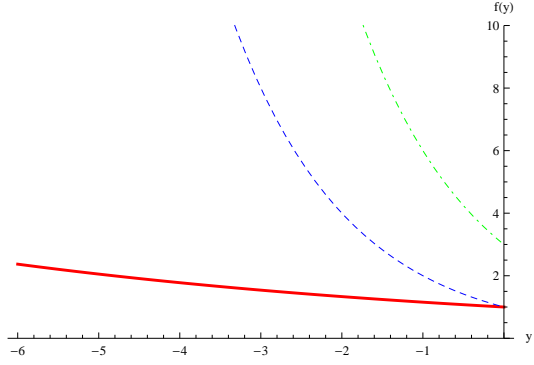}\\
  \caption{$f(y)$ is the value of $C^y_a(|\psi\rangle_{A|B_1B_2B_3})$. Solid (red) line is the exact value of $C^y_a(|\psi\rangle_{A|B_1B_2B_3})$, dashed (blue) line is the upper bound of $C^y_a(|\psi\rangle_{A|B_1B_2B_3})$ in (\ref{th3}), and dotdashed (green) line is the upper bound in \cite{ZXN}.}\label{2}
\end{figure}

{\it Remark 1}.~In (\ref{th3}) we have assumed that all $C(\rho_{AB_{j_i}})$, $i=1,2,\cdots,m-1$, are nonzero.
In fact, if one of them is zero, the inequality still holds by removing this term from the inequality. Namely, if $C(\rho_{AB_{j_i}})=0,$ then one has $C_a^y(\rho_{A|B_{j_1}\cdots B_{j_{m-1}}})<\frac{1}{2}C_a^y(\rho_{AB_{j_1}})+\cdots+\left(\frac{1}{2}\right)^{i-1}C_a^y(\rho_{AB_{j_{i-1}}})+\left(\frac{1}{2}\right)^{i}C_a^y(\rho_{AB_{j_{i+1}}})+\cdots+\left(\frac{1}{2}\right)^{m-3}C_a^y(\rho_{AB_{j_{m-2}}})+\left(\frac{1}{2}\right)^{m-3} C_a^y(\rho_{AB_{j_{m-1}}})$. By cyclically permuting the sub-indices in $B_{j_1}\cdots B_{j_{m-1}}$, we can get a set of inequalities. Summing up these inequalities we have  $C_a^y(\rho_{A|B_{j_1}\cdots B_{j_{m-1}}})<\frac{1}{m-1}\left(C_a^y(\rho_{AB_{j_1}})+\cdots+C_a^y(\rho_{AB_{j_{i-1}}})+C_a^y(\rho_{AB_{j_{i+1}}})+\cdots+C_a^y(\rho_{AB_{j_{m-2}}})+ C_a^y(\rho_{AB_{j_{m-1}}})\right)$ for $y<0$.

\section{MONOGAMY RELATIONS FOR NAGATIVITY OF ASSISTANCE}

Another well-known quantifier of bipartite entanglement is the negativity. Given a bipartite state $\rho_{AB}$ in $H_A\otimes H_B$, the negativity is defined by \cite{GRF},
$N(\rho_{AB})=(||\rho_{AB}^{T_A}||-1)/2$,
where $\rho_{AB}^{T_A}$ is the partial transpose with respect to the subsystem $A$, $||X||$ denotes the trace norm of $X$, i.e $||X||=\mathrm{Tr}\sqrt{XX^\dag}$.
Negativity is a computable measure of entanglement, and is a convex function of $\rho_{AB}$. It vanishes if and only if $\rho_{AB}$ is separable for the $2\otimes2$ and $2\otimes3$ systems \cite{MPR}. For the purpose of discussion, we use the following definition of negativity, $ N(\rho_{AB})=||\rho_{AB}^{T_A}||-1$.
For any bipartite pure state $|\psi\rangle_{AB}$, the negativity $ N(\rho_{AB})$ is given by
$N(|\psi\rangle_{AB})=2\sum_{i<j}\sqrt{\lambda_i\lambda_j}=(\mathrm{Tr}\sqrt{\rho_A})^2-1$,
where $\lambda_i$ are the eigenvalues for the reduced density matrix of $|\psi\rangle_{AB}$. For a mixed state $\rho_{AB}$, the convex-roof extended negativity (CREN) is defined as
\begin{equation}\label{nc}
 N_c(\rho_{AB})=\mathrm{min}\sum_ip_iN(|\psi_i\rangle_{AB}),
\end{equation}
where the minimum is taken over all possible pure state decompositions $\{p_i,~|\psi_i\rangle_{AB}\}$ of $\rho_{AB}$. CREN gives a perfect discrimination of positive partial transposed bound entangled states and separable states in any bipartite quantum systems \cite{PH,WJM}. For a mixed state $\rho_{AB}$, the
convex-roof extended negativity of assistance (CRENOA) is defined as \cite{JAB}
\begin{equation}\label{na}
 N_a(\rho_{AB})=\mathrm{max}\sum_ip_iN(|\psi_i\rangle_{AB}),
\end{equation}
where the maximum is taken over all possible pure state decompositions $\{p_i,~|\psi_i\rangle_{AB}\}$ of $\rho_{AB}$.

Let us consider the relation between CREN and concurrence. For any bipartite pure state    $|\psi\rangle_{AB}$ in a $d\otimes d$ quantum system with Schmidt rank 2,
$|\psi\rangle_{AB}=\sqrt{\lambda_0}|00\rangle+\sqrt{\lambda_1}|11\rangle$,
one has
$N(|\psi\rangle_{AB})=\parallel|\psi\rangle\langle\psi|^{T_B}\parallel-1=2\sqrt{\lambda_0\lambda_1}
=\sqrt{2(1-\mathrm{Tr}\rho_A^2)}=C(|\psi\rangle_{AB})$. In other words, negativity is equivalent to concurrence for any pure state with Schmidt rank 2, and consequently it follows that for any two-qubit mixed state $\rho_{AB}=\sum p_i|\psi_i\rangle_{AB}\langle\psi_i|$,
\begin{eqnarray}\label{N1}
 N_c(\rho_{AB})&&=\mathrm{min}\sum_ip_iN(|\psi_i\rangle_{AB})\\ \nonumber
&&=\mathrm{min}\sum_ip_iC(|\psi_i\rangle_{AB})\\ \nonumber
&&=C(\rho_{AB}),
\end{eqnarray}
\begin{eqnarray}\label{N2}
 N_a(\rho_{AB})&&=\mathrm{max}\sum_ip_iN(|\psi_i\rangle_{AB})\\ \nonumber
&&=\mathrm{max}\sum_ip_iC(|\psi_i\rangle_{AB})\\ \nonumber
&&= C_a(\rho_{AB}),
\end{eqnarray}
where the minimum and the maximum are taken over all pure state decompositions $\{p_i,~|\psi_i\rangle_{AB}\}$ of $\rho_{AB}$.

Combing (\ref{la2}), (\ref{N1}) and (\ref{N2}), we can get the following Lemma.

{\bf [Lemma 1].} For $N$-qubit generlized $W$-class states (\ref{gw}), we have
\begin{eqnarray}\label{la3}
N_c(\rho_{AB_i})=N_a(\rho_{AB_i}).
\end{eqnarray}

As is already known, the negativity satisfies the monogamy relation for N-qubit pure state \cite{JAB}. In fact, for any N-qubit state, the monogamy relation of the negativity always holds. Therefore, we can get the following Lemma.

{\bf [Lemma 2].} For any N-qubit state $\rho\in H_A\otimes H_{B_1}\otimes\cdots\otimes H_{B_{N-1}}$, we have
\begin{eqnarray}\label{la4}
N_c^x(\rho_{A|B_1\cdots B_{N-1}})\geq \sum_{i=1}^{N-1}N_c^x(\rho_{AB_i}),~~~~x\geq 2.
\end{eqnarray}

{\sf [Proof].} From Ref \cite{JAB}, one has
\begin{eqnarray}\label{lapf41}
N_c^2(|\psi\rangle_{A|B_1\cdots B_{N-1}})\geq \sum_{i=1}^{N-1}N_c^2(\rho_{AB_i}),
\end{eqnarray}
for N-qubit pure state. Applying the similar approach in Ref \cite{ZF}, one can get
\begin{eqnarray}\label{lapf42}
N_c^x(|\psi\rangle_{A|B_1\cdots B_{N-1}})\geq \sum_{i=1}^{N-1}N_c^x(\rho_{AB_i}),
\end{eqnarray}
for N-qubit pure state with $x\geq 2$.

Let $\rho=\sum_ip_i|\psi_i\rangle_{AB_1\cdots B_{N-1}}\langle \psi_i|$ be the optimal decomposition of $N_c(\rho_{A|B_1\cdots B_{N-1}})$ for the N-qubit mixed state, we have
\begin{eqnarray}\label{lapf43}
N_c^x(\rho_{A|B_1\cdots B_{N-1}})&&= \left(\sum_{i=1}p_iN_c(|\psi\rangle_{A|B_1\cdots B_{N-1}})\right)^x\\ \nonumber
&&\geq  \left(\sum_{i=1}p_i\sqrt{\sum_{k=1}^{N-1}N_c^2(\rho_{AB_k})}\right)^x\\ \nonumber
&&\geq \left[\sum_k\left(\sum_ip_iN_c(\rho_{AB_k})\right)^2\right]^\frac{x}{2}\\ \nonumber
&&\geq \sum_{i=1}^{N-1}N_c^x(\rho_{AB_i}),
\end{eqnarray}
where the first inequality is due to (\ref{lapf41}). The second inequality is due to Minkowski inequality: $(\sum_k(\sum_ix_{ik}))^\frac{1}{2}\leq \sum_i(\sum_kx_{ik}^2)^\frac{1}{2}$. The last inequality is due to $(\sum_ia_i)^\alpha\geq \sum_ia_i^\alpha$ for $a_i\geq 0,~\alpha\geq1$. \hfill \rule{1ex}{1ex}

In the following, we can derive a better monogamy relation for CREN.

{[\bf Lemma 3]}. For any N-qubit state $\rho\in H_A\otimes H_{B_1}\otimes\cdots\otimes H_{B_{N-1}}$, if
$N_c(\rho_{AB_i})\geq N_c(\rho_{A|B_{i+1}\cdots B_{N-1}})$ for $i=1, 2, \cdots, m$, and
$N_c(\rho_{AB_j})\leq N_c(\rho_{A|B_{j+1}\cdots B_{N-1}})$ for $j=m+1,\cdots,N-2$,
$\forall$ $1\leq m\leq N-3$, $N\geq 4$, we have
\begin{eqnarray}\label{la6}
&&N^x_c(\rho_{A|B_1B_2\cdots B_{N-1}})\geq N^x_c(\rho_{AB_1})\\\nonumber
 &&+\frac{x}{2} N^x_c(\rho_{AB_2})+\cdots+\left(\frac{x}{2}\right)^{m-1}N^x_c(\rho_{AB_m})\\\nonumber
 &&+\left(\frac{x}{2}\right)^{m+1}(N^x_c(\rho_{AB_{m+1}})
 +\cdots+N^x_c(\rho_{AB_{N-2}}))\\\nonumber
 &&+\left(\frac{x}{2}\right)^{m}N^x_c(\rho_{AB_{N-1}})
\end{eqnarray}
for all $x\geq2$.

{\sf [Proof].} From (\ref{la4}), one has $N^2_c(\rho_{A|BC})\geq N^2_c(\rho_{AB})+N^2_c(\rho_{AC}).$
If $N_c(\rho_{AB})\geq N_c(\rho_{AC})$, we have
\begin{eqnarray}\label{pfla61}
  N^x_c(\rho_{A|BC})&&\geq (N^2_c(\rho_{AB})+N^2_c(\rho_{AC}))^{\frac{x}{2}}=N^x_c(\rho_{AB})\left(1+\frac{N^2_c(\rho_{AC})}{N^2_c(\rho_{AB})}\right)^{\frac{x}{2}} \\\nonumber
   && \geq N^x_c(\rho_{AB})\left[1+\frac{x}{2}\left(\frac{N^2_c(\rho_{AC})}{N^2_c(\rho_{AB})}\right)^{\frac{x}{2}}\right]=N^x_c(\rho_{AB})+\frac{x}{2}N^x_c(\rho_{AC}),
\end{eqnarray}
where the second inequality is due to the inequality $(1+t)^x\geq 1+xt \geq 1+xt^x$ for $x\geq1,~0\leq t\leq1$.

By using the inequality (\ref{pfla61}) repeatedly, one gets
\begin{eqnarray}\label{pfla62}
&&N^x_c(\rho_{A|B_1B_2\cdots B_{N-1}})\geq  N^x_c(\rho_{AB_1})+\frac{x}{2}N^x_c(\rho_{A|B_2\cdots B_{N-1}})\\\nonumber
 &&\geq N^x_c(\rho_{AB_1})+\frac{x}{2}N^x_c(\rho_{AB_2})
 +\left(\frac{x}{2}\right)^2N^x_c(\rho_{A|B_3\cdots B_{N-1}})\\\nonumber
 &&
  \geq\cdots\geq N^x_c(\rho_{AB_1})+\frac{x}{2}N^x_c(\rho_{AB_2})
  +\cdots+\left(\frac{x}{2}\right)^{m-1}N^x_c(\rho_{AB_m})\\\nonumber
 &&
 +\left(\frac{x}{2}\right)^m N^x_c(\rho_{A|B_{m+1}\cdots B_{N-1}}).
\end{eqnarray}
As $N_c(\rho_{AB_j})\leq N_c(\rho_{A|B_{j+1}\cdots B_{N-1}})$ for $j=m+1,\cdots,N-2$, by (\ref{pfla61}) we get
\begin{eqnarray}\label{pfla63}
N^x_c(\rho_{A|B_{m+1}\cdots B_{N-1}})\geq \frac{x}{2}N^x_c(\rho_{AB_{m+1}})+N^x_c(\rho_{A|B_{m+2}\cdots B_{N-1}})\nonumber\\
\geq \frac{x}{2}(N^x_c(\rho_{AB_{m+1}})+\cdots+N^x_c(\rho_{AB_{N-2}}))+N^x_c(\rho_{AB_{N-1}}).
\end{eqnarray}
Combining (\ref{pfla62}) and (\ref{pfla63}), we have Lemma 3.
\hfill \rule{1ex}{1ex}

We can also derive a bound of $N_c^x(\rho_{A|B_1B_2\cdots B_{N-1}})$ for $x<0$.

{\bf [Lemma 4].} For any N-qubit state $\rho\in H_A\otimes H_{B_1}\otimes\cdots\otimes H_{B_{N-1}}$, we have
\begin{equation}\label{la5}
  N_c^x(\rho_{A|B_1B_2\cdots B_{N-1}})< M' \left(N_c^x(\rho_{AB_1})+N_c^x(\rho_{AB_2})+\cdots+N_c^x(\rho_{AB_{N-1}})\right)
\end{equation}
for all $x<0$, where $M'=\frac{1}{N-1}$.

{\sf [Proof].} For arbitrary tripartite state, from (\ref{la4}) we have
\begin{eqnarray}\label{lapf51}
  N^x_c(\rho_{A|B_1B_2})&&\leq \left(N_c^2(\rho_{AB_1})+N_c^2(\rho_{AB_2})\right)^{\frac{x}{2}}\\\nonumber
&&= N_c^x(\rho_{AB_1})\left(1+\frac{N_c^2(\rho_{AB_2})}{N_c^2(\rho_{AB_1})}\right)^{\frac{x}{2}}<N_c^x(\rho_{AB_1}),
\end{eqnarray}
where the first inequality is due to $x<0$ and the second inequality is due to $\left(1+\frac{N_c^2(\rho_{AB_2})}{N_c^2(\rho_{AB_1})}\right)^{\frac{x}{2}}<1.$
On the other hand, we have
\begin{eqnarray}\label{lapf52}
 N_c^x(\rho_{A|B_1B_2})&&\leq \left(N_c^2(\rho_{AB_1})+N_c^2(\rho_{AB_2})\right)^{\frac{x}{2}}\\\nonumber
 &&= N_c^x(\rho_{AB_2})\left(1+\frac{N_c^2(\rho_{AB_1})}{N_c^2(\rho_{AB_2})}\right)^{\frac{x}{2}}<N_c^x(\rho_{AB_2}).
\end{eqnarray}
From (\ref{lapf51}) and (\ref{lapf52}) we obtain
\begin{equation}\label{lapf53}
 N_c^x(\rho_{A|B_1B_2})< \frac{1}{2}(N_c^x(\rho_{AB_1})+N_c^x(\rho_{AB_2})).
\end{equation}
By using the inequality (\ref{lapf53}) repeatedly, one gets
\begin{eqnarray}\label{lapf54}
N_c^x(\rho_{A|B_1B_2\cdots B_{N-1}})&&< \frac{1}{2}\left(N_c^x(\rho_{AB_1})+N_c^x(\rho_{A|B_2\cdots B_{N-1}})\right)\\ \nonumber
&&<\frac{1}{2}N_c^x(\rho_{AB_1})+\left(\frac{1}{2}\right)^2N_c^x(\rho_{AB_2})
 +\left(\frac{1}{2}\right)^2N_c^x(\rho_{A|B_3\cdots B_{N-1}})\\ \nonumber
 &&<\cdots< \frac{1}{2}N_c^x(\rho_{AB_1})+\left(\frac{1}{2}\right)^2N_c^x(\rho_{AB_2})+\cdots\\ \nonumber
  &&+\left(\frac{1}{2}\right)^{N-2}N_c^x(\rho_{AB_{N-2}}) +\left(\frac{1}{2}\right)^{N-2} N_c^x(\rho_{AB_{N-1}}).
\end{eqnarray}

By cyclically permuting the sub-indices $B_1, B_2, \cdots, B_{N-1}$ in (\ref{lapf54}) we can get a set of inequalities. Summing up these inequalities we obtain (\ref{la5}). \hfill \rule{1ex}{1ex}

In the following, we study the monogamy property of the CRENOA for the $N$-qubit generalized $W$-class states (\ref{gw}). We can obtain the following theorem.

{[\bf Theorem 4]}.
For the $N$-qubit generalized $W$-class states $|\psi\rangle\in H_A\otimes H_{B_1}\otimes\cdots\otimes H_{B_{N-1}}$, with $\rho_{AB_{j_1}\cdots B_{j_{m-1}}}$ the $m$-qubit, $2\leq m\leq N$, reduced density matrix of $|\psi\rangle$. If $N_c(\rho_{AB_{j_i}})\geq N_c(\rho_{AB_{j_{i+1}}\cdots B_{j_{m-1}}})$ for $i=1,2,\cdots t$, and $N_c(\rho_{AB{j_k}})\leq N_c(\rho_{AB_{j_{k+1}}\cdots B_{j_{m-1}}})$ for $k=t+1,\cdots,{m-2}$, $\forall$ $1\leq t\leq {m-3},~m\geq 4$, then the CRENOA satisfies
\begin{eqnarray}\label{th4}
&&N_a^x(\rho_{A|B_{j_1}\cdots B_{j_{m-1}}})\geq N_a^x(\rho_{AB_{j_1}}) \nonumber\\
&&+\frac{x}{2}N_a^x(\rho_{AB_{j_2}})+\cdots+\left(\frac{x}{2}\right)^{t-1}N_a^x(\rho_{AB_{j_t}}) \nonumber\\
&&+\left(\frac{x}{2}\right)^{t+1}\left(N_a^x(\rho_{AB_{j_{t+1}}})+\cdots+N_a^x(\rho_{AB_{j_{m-2}}})\right)\nonumber\\
&&+\left(\frac{x}{2}\right)^tN_a^x(\rho_{AB_{j_{m-1}}})
\end{eqnarray}
for all $x\geq2$.

{\sf [Proof].}
For the $N$-qubit generalized $W$-class states $|\psi\rangle$, according to the definitions of $N_c(\rho)$ and $N_a(\rho)$, one has $N_a(\rho_{A|B_{j_1}\cdots B_{j_{m-1}}})\geq N_c(\rho_{A|B_{j_1}\cdots B_{j_{m-1}}})$. When $x\geq2$, we have
\begin{eqnarray}\label{pf4}
N_a^x(\rho_{A|B_{j_1}\cdots B_{j_{m-1}}})&&\geq N_c^x(\rho_{A|B_{j_1}\cdots B_{j_{m-1}}})
\geq N_c^x(\rho_{AB_{j_1}}) \nonumber\\
&&+\frac{x}{2}N_c^x(\rho_{AB_{j_2}})+\cdots+\left(\frac{x}{2}\right)^{t-1}N_c^x(\rho_{AB_{j_t}}) \nonumber\\
&&+\left(\frac{x}{2}\right)^{t+1}\left(N_c^x(\rho_{AB_{j_{t+1}}})+\cdots+N_c^x(\rho_{AB_{j_{m-2}}})\right)\nonumber\\
&&+\left(\frac{x}{2}\right)^t N_c^x(\rho_{AB_{j_{m-1}}})\nonumber\\
&&=N_a^x(\rho_{AB_{j_1}})
+\frac{x}{2}N_a^x(\rho_{AB_{j_2}})+\cdots+\left(\frac{x}{2}\right)^{t-1}N_a^x(\rho_{AB_{j_t}}) \nonumber\\
&&+\left(\frac{x}{2}\right)^{t+1}\left(N_a^x(\rho_{AB_{j_{t+1}}})+\cdots+N_a^x(\rho_{AB_{j_{m-2}}})\right)\nonumber\\
&&+\left(\frac{x}{2}\right)^tN_a^x(\rho_{AB_{j_{m-1}}}),
\end{eqnarray}
where we have used in the first inequality the relation $a^x\geq b^x$ for $a\geq b\geq 0,~x\geq 2$. Using the result of Lemma 3, one gets the second inequality. The equality is due to the Lemma 2.
\hfill \rule{1ex}{1ex}

In Theorem 4 we have assumed that
some $N_c(\rho_{AB_{j_i}})\geq N_c(\rho_{AB_{j_{i+1}}\cdots B_{j_{m-1}}})$ and some
$N_c(\rho_{AB_{j_k}})\leq N_c(\rho_{AB_{j_{k+1}}\cdots B_{j_{m-1}}})$ for the $N$-qubit generalized $W$-class states.
If all $N_c(\rho_{AB_{j_i}})\geq N_c(\rho_{AB_{j_{i+1}}\cdots B_{j_{m-1}}})$ for $i=1, 2, \cdots, {m-2}$, then we have
the following conclusion:

{\bf [Theorem 5]}.
If $N_c(\rho_{AB_{j_i}})\geq N_c(\rho_{AB_{j_{i+1}}\cdots B_{j_{m-1}}})$ for $i=1, 2, \cdots, {m-2}$, we have
\begin{eqnarray}\label{th5}
N_a^x(\rho_{A|B_{j_1}\cdots B_{j_{m-1}}})\geq N_a^x(\rho_{AB_{j_1}}) +\frac{x}{2}N_a^x(\rho_{AB_{j_2}})+\cdots+\left(\frac{x}{2}\right)^{m-2}N_a^x(\rho_{AB_{j_{m-1}}})
\end{eqnarray}
for all $x\geq2$.

We can also derive a tighter upper bound of $N_a^y(\rho_{AB_1\cdots B_{N-1}})$ for $y<0$.

{[\bf Theorem 6]}.
For the $N$-qubit generalized $W$-class states $|\psi\rangle\in H_A\otimes H_{B_1}\otimes\cdots\otimes H_{B_{N-1}}$ with $N_c(\rho_{AB_{j_i}})\neq 0$ for $1\leq i\leq m-1$, we have
\begin{eqnarray}\label{th6}
  N_a^y(\rho_{A|B_{j_1}\cdots B_{j_{m-1}}})< \tilde{M} \left(N_a^y(\rho_{AB_{j_1}})+N_a^y(\rho_{AB_{j_2}})+\cdots+N_a^y(\rho_{AB_{j_{m-1}}})\right)
\end{eqnarray}
for all $y<0$, where $\tilde{M}=\frac{1}{m-1}$.

{\sf [Proof].}
For $y< 0$, we have
\begin{eqnarray}\label{pf61}
  N_a^y(\rho_{A|B_{j_1}\cdots B_{j_{m-1}}})&&\leq N_c^y(\rho_{A|B_{j_1}\cdots B_{j_{m-1}}}) \nonumber\\
&& <\tilde{M} \left(N_c^y(\rho_{AB_{j_1}})+N_c^y(\rho_{AB_{j_2}})+\cdots+N_c^y(\rho_{AB_{j_{m-1}}})\right)\nonumber \\
&&= \tilde{M} \left(N_a^y(\rho_{AB_{j_1}})+N_a^y(\rho_{AB_{j_2}})+\cdots+N_a^y(\rho_{AB_{j_{m-1}}})\right),
\end{eqnarray}
where we have used in the first inequality the relation $a^x\leq b^x$ for $a\geq b\geq 0,~x\leq 0$. The second inequality is based on Lemma 4. The equality is due to the Lemma 2. \hfill \rule{1ex}{1ex}

{\it Remark 2}.~In (\ref{th6}) we have assumed that all $N_c(\rho_{AB_{j_i}})$, $i=1,2,\cdots,m-1$, are nonzero.
In fact, if one of them is zero, the inequality still holds if one simply removes this term from the inequality. Namely, if $N_c(\rho_{AB_{j_i}})=0,$ then one has $N_a^y(\rho_{A|B_{j_1}\cdots B_{j_{m-1}}})<\frac{1}{2}N_a^y(\rho_{AB_{j_1}})+\cdots+\left(\frac{1}{2}\right)^{i-1}N_a^y(\rho_{AB_{j_{i-1}}})+\left(\frac{1}{2}\right)^{i}N_a^y(\rho_{AB_{j_{i+1}}})+\cdots+\left(\frac{1}{2}\right)^{m-3}N_a^y(\rho_{AB_{j_{m-2}}})+\left(\frac{1}{2}\right)^{m-3} N_a^y(\rho_{AB_{j_{m-1}}})$. By cyclically permuting the sub-indices in $B_{j_1}\cdots B_{j_{m-1}}$, we can get a set of inequalities. Summing up these inequalities we have  $N_a^y(\rho_{A|B_{j_1}\cdots B_{j_{m-1}}})<\frac{1}{m-1}\left(N_a^y(\rho_{AB_{j_1}})+\cdots+N_a^y(\rho_{AB_{j_{i-1}}})+N_a^y(\rho_{AB_{j_{i+1}}})+\cdots+N_a^y(\rho_{AB_{j_{m-2}}})+ N_a^y(\rho_{AB_{j_{m-1}}})\right)$, for $y<0$.

\section{conclusion}
Entanglement monogamy is a fundamental property of multipartite entangled states. We have presented tighter monogamy inequalities for the $x$-power of concurrence of assistance $C_a^x(\rho_{A|B_{j_1}\cdots B_{j_{m-1}}})$ of the $m$-qubit reduced density matrices, $2\leq m \leq N$, for the $N$-qubit generalized $W$-class states, when $x\geq 2$. A tighter upper bound of $y$-power of concurrence of assistance is also derived for $y<0$. The monogamy relations for the $x$-power of negativity of assistance for the $N$-qubit generalized $W$-class states have been also investigated for $x\geq 2$ and $x<0$, respectively. These relations give rise to the restrictions of entanglement distribution among the qubits in generalized $W$-class states.
It should be noted that entanglement of assistances like concurrence of assistance and negativity of assistance are not genuine measures of quantum entanglement.
They quantify the maximum average amount of entanglement between two parties, Alice and Bob, which can be extracted
given assistance from a third party, Charlie, by performing a measurement on his system and reporting the measurement outcomes to Alice and Bob.
Nevertheless, similar to quantum entanglement, we see that the entanglement of assistances also satisfy certain monogamy relations.

\bigskip
\noindent{\bf Acknowledgments}\, \, This work is supported by the NSF of China under Grant No. 11675113.

\end{document}